# Steering alkyne homocoupling with on-surface synthesized catalysts

Mohammed S. G. Mohammed [a,b], Luciano Colazzo [a,b,†], Aurelio Gallardo [c,d,g], José A. Pomposo [b,e,f], Pavel Jelínek [c,g], and Dimas G. de Oteyza [a,b,e,*]

[a.] Donostia International Physics Center (DIPC), 20018 San Sebastián, Spain.
[b.] Centro de Física de Materiales (CFM-MPC), CSIC-UPV/EHU, 20018 San Sebastián, Spain.
[c.] Institute of Physics, The Czech Academy of Sciences, 162 00 Prague, Czech Republic.
[d.] Faculty of Mathematics and Physics, Charles University, 180 00 Prague, Czech Republic.
[e.] Ikerbasque, Basque Foundation for Science, 48013 Bilbao, Spain.
[f.] Departamento de Física de Materiales, Universidad del País Vasco (UPV/EHU), Apartado 1072, E-20800 San Sebastián, Spain
[g.] RCPTM Palacký University Olomouc, 771 46 Olomouc, Czech Republic
* email: d_g_oteyza@ehu.es
† Present addresses: Center for Quantum Nanoscience, Institute for Basic Science (IBS), Seoul 03760, Republic of Korea and Department of Physics, Ewha Womans University, Seoul 03760, Republic of Korea.

**We report a multi-step on-surface synthesis strategy. The first step consists in the surface-supported synthesis of metal-organic complexes, which are subsequently used as catalysts to steer on-surface alkyne coupling reactions. In addition, we analyze and compare the electronic properties of the different coupling motifs obtained.**

Metal-organic complexes are a class of compounds formed by organic building units linked by coordination to metal ions. These complexes are shown to be extremely versatile systems for many applications that include e.g. from biological and medical use[1–4] to hydrogen storage,[5,6] $CO_2$ sequestration,[7] filtering [8] or catalysis.[9,10] Their functionality frequently relies on the properties of the ligand and/or on their structure, which often forms long-range ordered porous frameworks (so-called metal-organic frameworks or MOFs). However, the catalytic activity is frequently determined by the metal centers.

Despite the booming development of "on-surface synthesis" strategies, whereby chemical reactions are driven under the confinement of well-defined surfaces (often under vacuum conditions),[11] the catalytic use of 2D metal-organic complexes to influence such surface-supported reactions has hardly been explored to date. In this work we will study the catalytic effect of Au-thiolate structures on one of the most popular reaction schemes applied in on-surface synthesis: alkyne homocoupling.[12] This is a particularly relevant reaction, among other reasons, because it maintains the conjugation between the reactant units and can thus be used in the synthesis of functional organic semiconductors.[12]

In a previous work we reported how to control the stereospecific bonding motif in the formation of Au-thiolate links. Thereby, we could controllably form triangular Au-coordinated metal-organic complexes with 1,4-bis(4-mercaptophenyl)benzene (BMB) as organic ligands. [13] Now, we study their catalytic effect on the dimerization of 1-ethynyl-pyrene reactants and further compare the electronic properties of the different coupling schemes.

Upon deposition of 1-ethynyl-pyrene (**m1**, Fig. 1a) onto Au(111) held at room temperature (RT) the molecules remain unreacted (Fig.1b) and are found either as monomers (Fig. 1c) or as non-covalently bound dimers (Fig. 1d). The latter are clearly dominant, evidencing attractive intermolecular interactions that, however, do not drive formation of larger clusters. The dimers preferentially adsorb along the fcc trenches of the Au(111) herringbone reconstruction and display a clearly correlated inter-spacing along that one-dimensional confinement. These findings are all reminiscent of those found with other comparably sized molecular systems, in which the dimerization was associated to attractive dipole-dipole interactions [14,15] and the correlated inter-dimer spacing to electrostatic repulsion.[14,16]

As we anneal the sample to 470 K, the molecules react and couple covalently. Given the pronounced reactivity of the ethynyl group, the various types of reaction products all arise from alkyne homocoupling processes that result either in covalently bound dimers or trimers (Fig. 1e). Henceforth we will focus on the dimer structures.

As known from previous reports, alkyne homocoupling can result in a variety of different coupling schemes.[12,17] A particularly successful technique to determine the adsorbate's covalent bonding structures at the single molecule level is scanning probe microscopy with CO-functionalized tips in the repulsive tip-sample interaction regime.[18–20] Applying this imaging technique to the dominant dimer product in constant height scanning tunneling microscopy (STM) mode renders images as shown in Fig.1g. From the non-linear arrangement of the "inter-pyrene" bonding motif and the lack of a mirror symmetry plane for the resulting product structure, dehydrogenative Glaser coupling can be discarded. Instead, the products can be assigned to structure **d1** (Fig. 1e), further supported by comparison to simulations with the particle probe model (PPM, Fig. 1h).[20,21] This structure stems from a non-dehydrogenative alkyne coupling reaction[12,17,22] and, surprisingly, all covalently coupled dimers share the same linking structure, whether in cis or trans configuration (Fig. S1). Note that, as opposed to non-contact atomic force microscopy (nc-AFM) imaging, in which the higher electron density at triple bonds makes them appear with higher contrast,[19,20,23] in STM mode they appear as a nodal plane in the current signal (Fig.

1g),[23] more obvious in the experimental data than in the PPM simulations.

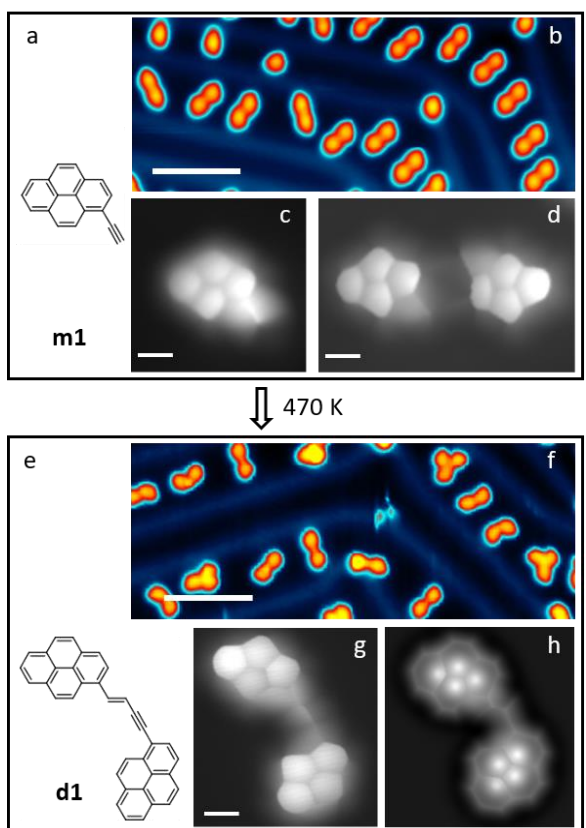

**Fig. 1.** Visualization of conventional on-surface alkyne homocoupling on Au(111) a) Chemical structure of the **m1** precursor. b) STM overview of the sample after **m1** deposition on Au(111) held at RT. c) High-resolution STM image with CO-functionalized tips of the monomer. d) Similar image of a non-covalently bound dimer. e) Chemical structure of **d1** product. f) Overview image of the sample after annealing to 470 K. g) High-resolution STM image with CO-functionalized tips of **d1**. h) Particle probe model image simulation of **d1**. d) Similar image of a non-covalently bound dimer. Imaging parameters and scalebars: b) U= 1.0 V/I= 100 pA, 5 nm c) U= 2 mV, 3 Å d) U= 2 mV, 3 Å f) U= −0.5 V/I= 10 pA, 5 nm g) U= 2 mV, 3 Å.

In the following we describe the changes in the reactivity of **m1** brought about by the presence of Au-thiolate-based metal-organic complexes. The $Au_3BMB_3$ complexes (Fig. 2b,c) are first formed, as described earlier,[13] by deposition of BMB (Fig. 2a) on Au(111) held at RT. Subsequently, the reactant **m1** is deposited and the sample is annealed. A representative image of the sample in vicinity of a $Au_3BMB_3$ complex after annealing to 360 K is shown in Fig. 2d. Whereas further from metal-organic complexes **m1** remains mostly unreacted (as illustrated e.g. by all the monomers and dimers in Fig. 2d that are not next to the metal-organic complex), near $Au_3BMB_3$ the scenario is completely different, providing evidence for their catalytic effect by reducing the threshold temperature for their covalent coupling reaction.

A standing out observation is the remarkable tendency of the Au-thiolate vertices of the $Au_3BMB_3$ triangles to become decorated by pyrene dimers. High-resolution imaging with CO-functionalized probes, however, is greatly distorted by the proximity of the metal-organic vertex (Fig. 2e) and does not allow assigning a particular molecular structure to those dimers. As the distorting interactions of the CO with the metal-organic vertex are avoided by manipulating the dimers away from it, the high resolution imaging (Fig. 2f) eventually allows for the determination of the dimer's structure **d2** (Fig. 2g, note again the nodal plane in the current signal at the triple bond position), further confirmed with PPM simulations (Fig. 2h) that nicely match the experimental data.

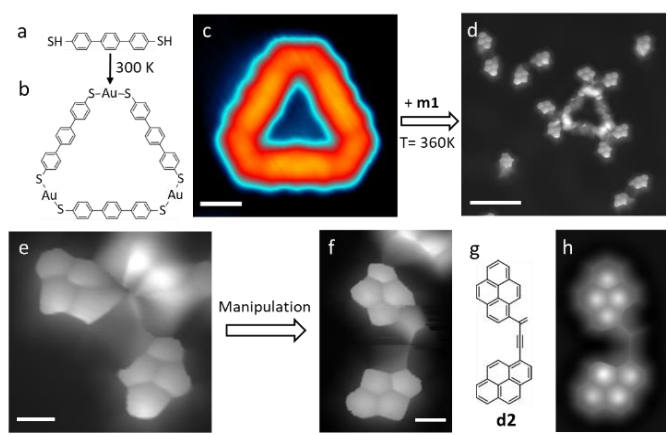

**Fig. 2.** Steering alkyne homocoupling with on-surface synthesized metal-organic complexes. a) Chemical structure of the BMB molecules and b) chemical structure of the $Au_3BMB_3$ complex formed on Au(111). c) Constant-height STM image of a $Au_3BMB_3$ and **m1** sample after annealing to 360 K. e) High-resolution constant-current image the **d2** dimer next to the metal-organic vertex and f) after manipulating **d2** away from it. g) Chemical structure of **d2**. h) Particle probe model simulation of **d2**. Imaging parameters and scale bars: c) U= 150 mV/I= 500 pA, SB= 6 Å d) U= 2 mV, 2 nm e) U= 2 mV, 3 Å f) U= 2 mV, 3 Å.

However, one has to be sure that this manipulation process does not change the dimer's structure. To check this, we perform scanning tunneling spectroscopy (STS) measurements on the **d2** dimers before and after manipulation (Fig. S2). The unaffected dI/dV signal, which is proportional to the local density of states, provides evidence for the unchanged chemical structure. This is further confirmed by the reversibility of the manipulation: the attractive interactions between organo-metallic vertex and **d2** often cause a spontaneous return of the manipulated dimer back to the vertex, resulting in a configuration that is indistinguishable from the starting point (Fig. S3).

It is worth noting that, whereas only one type of covalent coupling motif was found for the dimers on Au(111) in the absence of the metal-organic complexes, only 76% of the dimers at the $Au_3BMB_3$ vertices are **d2**. The remaining 24% are **d1** (no trimers are found next to the vertices). It remains unclear, however, whether the **d1** dimers also form next to the metal-organic vertices. Although only in low numbers, some covalently coupled dimers (**d1**) and trimers are also found on the Au(111) surface farther from the $Au_3BMB_3$ complexes upon

360 K annealing. Thus, we cannot discard that **d1** is formed on the bare Au(111) surface and only later diffuses to the metal-organic vertices, which could in turn imply that the selectivity for **d2** at the Au$_3$BMB$_3$ vertices is close to 100 %. Not being able to discern the two scenarios from our experiments, we can only claim that the 76% yield for **d2** at the metal-organic vertices is a lower limit value.

The two types of dimers observed are not only different in the imaging, but also in the electronic properties. This is an important point, since, apart from Glaser-type coupling,[12,17] these are the other two possible linkage schemes that may result from linear alkyne coupling reactions (besides the branched 2D structures that are obtained if a third ethynyl group is involved).[12,17] The electronic properties of organic materials greatly depend on their connectivity.[24] Relevant examples thereof are linear on-surface synthesized polymers whose bandgap can be tuned as a function of the linking bridges between periodic polyaromatic hydrocarbon units.[25,26] Indeed, an appropriate linkage between periodic pyrene units (as used in this work) has been shown to allow for the creation of remarkably low bandgap polymers.[25]

In general, the bandgap of a material decreases as the electron delocalization increases.[24,27] Applied to our dimers, the better the electronic coupling between the two pyrene units, the more delocalized will be the electrons over the whole dimer structure, which will consequently be mirrored in a lower bandgap. Representative spectra for **m1**, **d1** and **d2** are shown in Fig. 3a. As expected, the monomer displays the largest bandgap. The second largest bandgap value is found for **d2**, with only a minor decrease from 3.8 eV to 3.4 eV. A somewhat larger decrease down to 2.8 eV is found for **d1**. The difference between **d1** and **d2** can be rationalized as follows.

Whereas **d1** is conjugated across the whole dimer structure, in **d2** one of the double-bonds in the inter-pyrene linkage branches off rather than continuing along the linear coupling motif and causes its cross-conjugation.[27–29] That is, each of the pyrene units extends its conjugation up the branching point, but not with one another. This effect reduces the electron delocalization and causes a larger bandgap closer to that of the monomer.[27,28]

Evidence for the different bandgaps can also be obtained from dI/dV maps of Au$_3$BMB$_3$ complexes decorated with the two types of dimers (Fig. 3b-d, with **d1** on the upper-right vertex and **d2** on the other two vertices). The lowest unoccupied molecular orbitals (LUMO) of **d1** and **d2** only show a minor energy shift, much smaller than their considerable width. That is, the LUMO levels largely overlap and conductance maps around the LUMO energy thus show intensity on both kinds of dimers, as shown in Fig. 3d at 1.8 eV. The highest occupied molecular orbitals (HOMO), however, are not only sharper but also show a larger energy shift, making it clearly recognizable also in conductance maps. As such, a conductance map at -1.0 eV clearly shows stronger intensity on the lower bandgap structure **d1** (Fig. 3c), whereas at lower energy (e.g. at -1.4 eV in Fig. 3b) the signal becomes dominant on **d2**.

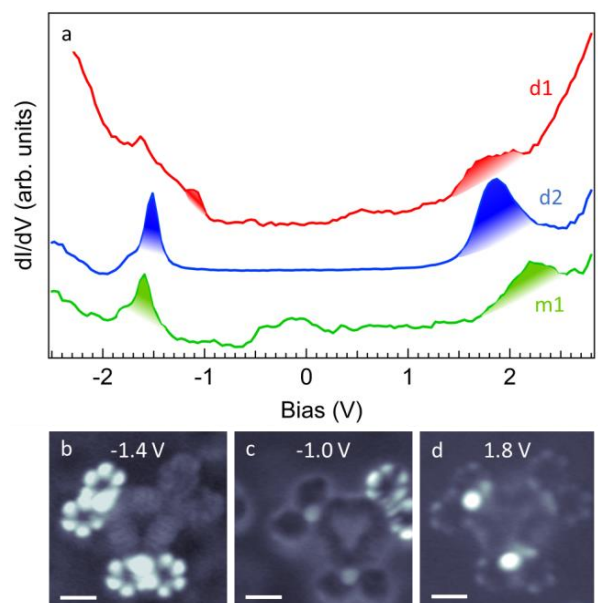

**Fig. 3.** a) Comparative dI/dV spectra taken on a monomer (green line), a **d1** dimer (red line) and a **d2** dimer (blue line). The energies of the HOMO and LUMO are shown by shadow colors in corresponding spectra b-d) Constant-current dI/dV maps of a Au$_3$BMB$_3$ complex decorated with **d1** on the upper-right vertex and **d2** on the other two vertices at the HOMO energy of **d2** b), at the HOMO energy of **d1** c) and at the LUMO energy of both d). Scale bars: 1 nm.

Altogether, in this work we have taken a step further in the development of on-surface synthesis. First, we proceed with the surface-supported synthesis of metal-organic complexes that are then used as catalysts to steer on-surface alkyne coupling reactions. In doing that, the threshold temperature for the reaction activation is lowered and the dominant product outcome is modified. Whereas in the absence of the catalyst a conjugated link is formed between the pyrene units that fosters the electron delocalization and lowers the overall structure's bandgap, the catalyst promotes the formation of a cross-conjugated link with a comparatively lower electron delocalization and thus with a larger bandgap.

Acknowledgements: The project leading to this publication has received funding from the European Research Council (ERC) under the European Union's Horizon 2020 research and innovation programme (grant agreement No 635919) and from the Spanish Ministry of Economy, Industry and Competitiveness (MINECO, Grant No. MAT2016-78293-C6-1-R). P. J. and A. G. acknowledge the support from Praemium Academie of the Academy of Science of the Czech Republic and GACR 18-09914S.

## Notes and references

# Steering alkyne homocoupling with on-surface synthesized catalysts

Mohammed S. G. Mohammed [a,b], Luciano Colazzo [a,b, ‡], Aurelio Gallardo [c,d,g], José A. Pomposo [b,e,f], Pavel Jelínek [c,g], and Dimas G. de Oteyza [a,b,e]

[a] *Donostia International Physics Center (DIPC), 20018 San Sebastián, Spain.*

[b] *Centro de Física de Materiales (CFM-MPC), CSIC-UPV/EHU, 20018 San Sebastián, Spain.*

[c] *Institute of Physics, The Czech Academy of Sciences, 162 00 Prague, Czech Republic.*

[d] *Faculty of Mathematics and Physics, Charles University, 180 00 Prague, Czech Republic.*

[e] *Ikerbasque, Basque Foundation for Science, 48013 Bilbao, Spain*

[f] *Departamento de Física de Materiales, Universidad del País Vasco (UPV/EHU), Apartado 1072, E-20800 San Sebastián, Spain*

[g] *RCPTM Palacky University Olomouc, 771 46 Olomouc, Czech Republic.*

‡ *Present addresses: Center for Quantum Nanoscience, Institute for Basic Science (IBS), Seoul 03760, Republic of Korea and Department of Physics, Ewha Womans University, Seoul 03760, Republic of Korea.*

METHODS

The experiments in this work were run on a commercial Scienta-Omicron low-temperature scanning probe system operating at 4.3 K under ultra-high vacuum. The surface of Au(111) single crystal of was cleaned by $Ar^+$ sputtering and subsequent annealing cycles. The crystal was then let to cool down to room temperature in the preparation chamber before the molecules were sublimed from home-built Knudsen-cell evaporators. The sublimation temperatures of BMB and 1-ethynyl pyrene molecules are 380 K and 300 K respectively.

STM measurements were performed with a Pt/Ir tip sharpened by poking into the bare metal surface. For CO-functionalization of the STM tip, NaCl was deposited on the surface by sublimed at 780 K. The sample is then moved to the STM, cooled down to 4.3 K and exposed to a CO partial pressure of 5E-9 mBar in the

STM chamber with open LT-STM shields. The gas is introduced through a leak valve in the preparation chamber, for which the gate valve between the chambers is opened. The maximum sample temperature reached during this CO deposition process is 7 K. The only reason for NaCl deposition in this work is to facilitate the visualization of carbon monoxide molecules for the subsequent tip functionalization, which results in a notable resolution enhancement and allows for bond-resolving imaging of the molecular structures. The bias values indicated refer to sample bias with respect to the STM tip. dI/dV spectroscopy and mapping were performed with a lock-in amplifier with oscillation frequency set to 731 Hz. STM images were analyzed using WSxM software.[1]

Theoretical AFM images were calculated using the probe particle model.[2] The parameters of the tip were selected to mimic a CO-tip, using a quadrupole with a stiffness of 0.24 Nm-1 and a charge factor of -0.2 e. The electrostatic force was included in the AFM calculations using the hartree potential calculated by DFT.

DFT calculations were performed using the Fritz Haber Institute ab initio molecular simulations package (FHI-AIMS).[2] We used the general gradient approximation PBE potential and Van der Waals corrections were described by Tkatchenko–Scheffler method.[3,4] The structure was relaxed in a 1 layer gold slab with a energy convergence criteria of $10^{-5}$ eV.

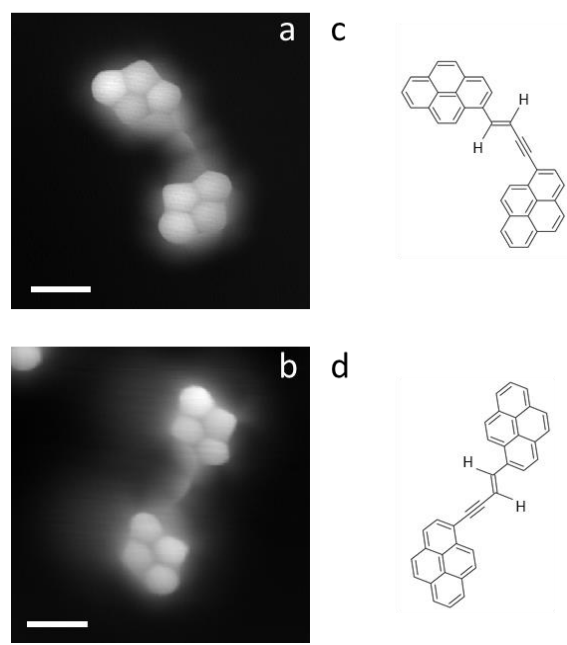

**Figure S1.** a-b) High-resolution STM images with CO-functionalized probes depicting the cis and trans dimer structures c-d) sharing the same coupling motif. For both a and b: U= 2 mV and Scalebars = 5 Å.

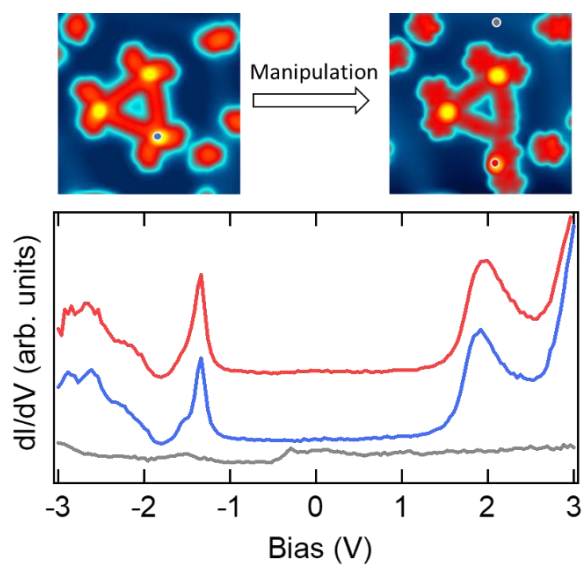

**Figure S2.** STM images of a **d2**-decorated Au$_3$BMB$_3$ complex before and after manipulation. The graph below shows the associated STS spectra of the **d2** dimer before (blue) and after (red) manipulation taken at the same position marked by blue and red dots on the images. The reference spectrum on the bare Au(111) surface is shown in grey.

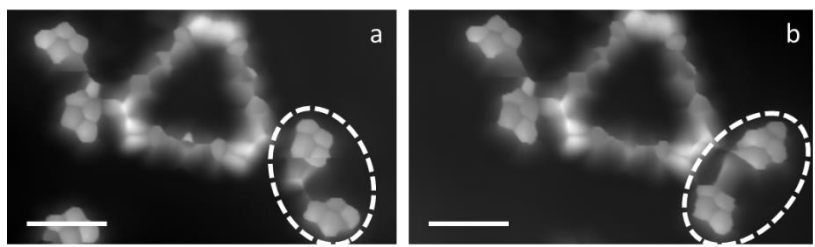

**Figure S3.** High-resolution imaging of **d2** dimer after manipulation (a) helps for better seeing the chemical structure which cannot be nicely resolved at the vertex due to the nonplanarity of the BMB triangles. By normal constant-height scanning as well as constant-current at low bias voltages we notice the tendency of the manipulated dimer to jump back to their initial position at the vertex (b) although STS data with the associated dI/dV maps show no strong bonding of the dimers at the vertices. Scalebar: 1nm for a and b. U = 2 mV.